\PassOptionsToPackage{usenames,dvipsnames}{xcolor}
\documentclass[modern]{descnote}
\pdfoutput=1 
\usepackage[T1]{fontenc}
\usepackage{ae,aecompl}
\usepackage[utf8]{inputenc}
\usepackage{newtxtext,newtxmath}
\usepackage[english]{babel}
\usepackage{amsmath,amstext}
\usepackage[figure,figure*]{hypcap}
\usepackage[hang]{footmisc}
\usepackage{threeparttablex}
\usepackage{longtable}
\usepackage{inconsolata}
\usepackage[frozencache,cachedir={_minted}]{minted}

\definecolor{DESCred}{rgb}{0.63,0.00,0.20}
\hypersetup{colorlinks=true,breaklinks=true,
  citecolor=DESCred,filecolor=DESCred,linkcolor=DESCred,urlcolor=DESCred}

\graphicspath{{figs/}}

\newcommand*{\https}[1]{\href{https://#1}{\nolinkurl{#1}}}
\newcommand*{\http}[1]{\href{http://#1}{\nolinkurl{#1}}}

\DeclareUrlCommand\code{\urlstyle{tt}}

\newcommand*{\thisversion}{v2}

\shorttitle{DESC DC2 Data Release Note}
\shortauthors{LSST~DESC}

\begin{document}
\title{DESC DC2 Data Release Note}
\collaboration{56}{The LSST Dark Energy Science Collaboration (LSST DESC)}

\author[0000-0003-1820-8486]{Bela Abolfathi}
\affiliation{Department of Physics and Astronomy, University of California, Irvine, Irvine, CA 92697, USA}

\author[0000-0002-6911-1038]{Robert Armstrong}
\affiliation{Lawrence Livermore National Laboratory, Livermore, CA 94550, USA}

\author[0000-0003-2296-7717]{Humna Awan}
\affiliation{Department of Physics and Astronomy, Rutgers, The State University of New Jersey, Piscataway, NJ 08854, USA}
\affiliation{Leinweber Center for Theoretical Physics, Department of Physics, University of Michigan, Ann Arbor, MI 48109, USA}

\author[0000-0002-9162-6003]{Yadu N. Babuji}
\affiliation{Argonne National Laboratory, Lemont, IL 60439, USA}
\affiliation{University of Chicago, Chicago, IL 60637, USA}

\author[0000-0002-8686-8737]{Franz Erik Bauer}
\affiliation{Instituto de Astrof{\'{\i}}sica and Centro de Astroingenier{\'{\i}}a, Facultad de F{\'{i}}sica, Pontificia Universidad Cat{\'{o}}lica de Chile, Casilla 306, Santiago 22, Chile} 
\affiliation{Millennium Institute of Astrophysics (MAS), Nuncio Monse{\~{n}}or S{\'{o}}tero Sanz 100, Providencia, Santiago, Chile} 
\affiliation{Space Science Institute, 4750 Walnut Street, Suite 205, Boulder, Colorado 80301}

\author[0000-0003-3623-9753]{George Beckett}
\affiliation{EPCC, University of Edinburgh, United Kingdom}

\author[0000-0002-5741-7195]{Rahul Biswas}
\affiliation{The Oskar Klein Centre for Cosmoparticle Physics, Stockholm University, AlbaNova, Stockholm, SE-106 91, Sweden}

\author[0000-0002-1345-1359]{Joanne R. Bogart}
\affiliation{SLAC National Accelerator Laboratory, Menlo Park, CA 94025, USA}
\affiliation{Kavli Institute for Particle Astrophysics and Cosmology, Stanford University, Stanford, CA  94305, USA}

\author[0000-0003-4887-2150]{Dominique Boutigny}
\affiliation{Univ. Grenoble Alpes, Univ. Savoie Mont Blanc, CNRS, LAPP, 74000 Annecy, France}

\author[0000-0002-7370-4805]{Kyle Chard}
\affiliation{Argonne National Laboratory, Lemont, IL 60439, USA}
\affiliation{University of Chicago, Chicago, IL 60637, USA}

\author[0000-0001-5738-8956]{James Chiang}
\affiliation{SLAC National Accelerator Laboratory, Menlo Park, CA 94025, USA}
\affiliation{Kavli Institute for Particle Astrophysics and Cosmology, Stanford University, Stanford, CA  94305, USA}

\author[0000-0001-9022-4232]{Johann Cohen-Tanugi}
\affiliation{Univ. Montpellier, CNRS, LUPM, 34095 Montpellier, France}
\affiliation{LPC, IN2P3/CNRS, Université Clermont Auvergne, F-63000 Clermont-Ferrand, France}

\author[0000-0001-5576-8189]{Andrew J. Connolly}
\affiliation{DIRAC Institute and Department of Astronomy, University of Washington, Seattle, WA 98195, USA}

\author{Scott F. Daniel}
\affiliation{DIRAC Institute and Department of Astronomy, University of Washington, Seattle, WA 98195, USA}

\author[0000-0002-5296-4720]{Seth W. Digel}
\affiliation{SLAC National Accelerator Laboratory, Menlo Park, CA 94025, USA}
\affiliation{Kavli Institute for Particle Astrophysics and Cosmology, Stanford University, Stanford, CA  94305, USA}

\author[0000-0001-8251-933X]{Alex Drlica-Wagner}
\affiliation{Fermi National Accelerator Laboratory, PO Box 500, Batavia, IL 60510, USA}
\affiliation{University of Chicago, Chicago IL 60637, USA}
\affiliation{Kavli Institute for Cosmological Physics, University of Chicago, Chicago, IL 60637, USA}

\author{Richard Dubois}
\affiliation{SLAC National Accelerator Laboratory, Menlo Park, CA 94025, USA}
\affiliation{Kavli Institute for Particle Astrophysics and Cosmology, Stanford University, Stanford, CA  94305, USA}

\author[0000-0003-1530-8713]{Eric Gawiser}
\affiliation{Department of Physics and Astronomy, Rutgers, The State University of New Jersey, Piscataway, NJ 08854, USA}

\author[0000-0001-9649-3871]{Thomas Glanzman}
\affiliation{SLAC National Accelerator Laboratory, Menlo Park, CA 94025, USA}
\affiliation{Kavli Institute for Particle Astrophysics and Cosmology, Stanford University, Stanford, CA  94305, USA}

\author[0000-0002-7832-0771]{Salman~Habib}
\affiliation{Argonne National Laboratory, Lemont, IL 60439, USA}

\author[0000-0003-2219-6852]{Andrew P.~Hearin}
\affiliation{Argonne National Laboratory, Lemont, IL 60439, USA}

\author[0000-0003-1468-8232]{Katrin~Heitmann}
\affiliation{Argonne National Laboratory, Lemont, IL 60439, USA}

\author[0000-0001-7203-2552]{Fabio Hernandez}
\affiliation{CNRS, CC-IN2P3, 21 avenue Pierre de Coubertin CS70202, 69627 Villeurbanne cedex, France}

\author[0000-0002-0965-7864]{Ren\'ee~Hlo\v{z}ek}
\affiliation{David A. Dunlap Department of Astronomy and Astrophysics, 50 St. George Street, Toronto ON M5S3H4}
\affiliation{Dunlap Institute for Astronomy and Astrophysics, 50 St. George Street, Toronto ON M5S3H4}

\author[0000-0002-8658-1672]{Joseph Hollowed}
\affiliation{Argonne National Laboratory, Lemont, IL 60439, USA}

\author[0000-0002-4179-5175]{Mike Jarvis}
\affiliation{Department of Physics \& Astronomy, University of Pennsylvania, 209 South 33rd Street, Philadelphia, PA 19104-6396, USA}

\author[0000-0001-8738-6011]{Saurabh~W.~Jha}
\affiliation{Department of Physics and Astronomy, Rutgers, The State University of New Jersey, Piscataway, NJ 08854, USA}

\author[0000-0002-6825-5283]{J. Bryce Kalmbach}
\affiliation{DIRAC Institute and Department of Astronomy, University of Washington, Seattle, WA 98195, USA}

\author[0000-0002-4394-6192]{Heather M. Kelly}
\affiliation{SLAC National Accelerator Laboratory, Menlo Park, CA 94025, USA}
\affiliation{Kavli Institute for Particle Astrophysics and Cosmology, Stanford University, Stanford, CA  94305, USA}

\author[0000-0002-2545-1989]{Eve Kovacs}
\affiliation{Argonne National Laboratory, Lemont, IL 60439, USA}

\author[0000-0003-0801-8339]{Danila Korytov}
\affiliation{Argonne National Laboratory, Lemont, IL 60439, USA}
\affiliation{University of Chicago, Chicago, IL 60637, USA}

\author[0000-0002-4410-7868]{K. Simon Krughoff}
\affiliation{Rubin Observatory Project Office, 950 N. Cherry Ave., Tucson, AZ 85719, USA}

\author{Craig S. Lage}
\affiliation{University of California-Davis, Davis, CA 95616, USA}

\author[0000-0001-7956-0542]{Fran\c{c}ois Lanusse}
\affiliation{AIM, CEA, CNRS, Universit\'e Paris-Saclay, Universit\'e Paris Diderot, Sorbonne Paris Cit\'e, F-91191 Gif-sur-Yvette, France}

\author[0000-0001-9592-4676]{Patricia Larsen}
\affiliation{Argonne National Laboratory, Lemont, IL 60439, USA}

\author[0000-0001-6800-7389]{Nan Li}
\affiliation{School of Physics and Astronomy, University of Nottingham, University Park, Nottingham, NG7 2RD, United Kingdom}

\author[0000-0002-6758-558X]{Emily Phillips Longley}
\affiliation{Department of Physics, Duke University, Durham NC 27708, USA}

\author[0000-0003-1666-0962]{Robert H. Lupton}
\affiliation{Princeton University, Princeton, NJ, USA}

\author[0000-0003-2271-1527]{Rachel Mandelbaum}
\affiliation{McWilliams Center for Cosmology, Department of Physics, Carnegie Mellon University, Pittsburgh, PA 15213, USA}

\author[0000-0002-1200-0820]{Yao-Yuan~Mao}
\affiliation{Department of Physics and Astronomy, Rutgers, The State University of New Jersey, Piscataway, NJ 08854, USA}
\affiliation{NASA Einstein Fellow}

\author[0000-0002-0113-5770]{Phil Marshall}
\affiliation{SLAC National Accelerator Laboratory, Menlo Park, CA 94025, USA}
\affiliation{Kavli Institute for Particle Astrophysics and Cosmology, Stanford University, Stanford, CA  94305, USA}

\author[0000-0002-2308-4230]{Joshua~E.~Meyers}
\affiliation{Lawrence Livermore National Laboratory, Livermore, CA 94550, USA}

\author[0000-0002-0692-1092]{Ji Won Park}
\affiliation{SLAC National Accelerator Laboratory, Menlo Park, CA 94025, USA}
\affiliation{Department of Physics, Stanford University, Stanford, CA 94305, USA}
\affiliation{Kavli Institute for Particle Astrophysics and Cosmology, Stanford University, Stanford, CA  94305, USA}

\author[0000-0002-8560-4449]{Julien Peloton}
\affiliation{Universit{\'e} Paris-Saclay, CNRS/IN2P3, IJCLab, Orsay, France}

\author[0000-0002-3988-4881]{Daniel Perrefort}
\affiliation{Department of Physics and Astronomy, University of Pittsburgh, Pittsburgh, PA 15260, USA}
\affiliation{Pittsburgh Particle Physics, Astrophysics and Cosmology Center (PITT PACC), University of Pittsburgh, Pittsburgh, PA 15260, USA}

\author{James Perry}
\affiliation{EPCC, University of Edinburgh, United Kingdom}

\author[0000-0002-1278-109X]{St\'ephane Plaszczynski}
\affiliation{Universit{\'e} Paris-Saclay, CNRS/IN2P3, IJCLab, Orsay, France}

\author[0000-0003-2265-5262]{Adrian~Pope}
\affiliation{Argonne National Laboratory, Lemont, IL 60439, USA}

\author[0000-0001-9376-3135]{Eli~S.~Rykoff}
\affiliation{SLAC National Accelerator Laboratory, Menlo Park, CA 94025, USA}
\affiliation{Kavli Institute for Particle Astrophysics and Cosmology, Stanford University, Stanford, CA  94305, USA}

\author[0000-0003-3136-9532]{F. Javier S\'{a}nchez}
\affiliation{Department of Physics and Astronomy, University of California, Irvine, Irvine, CA 92697, USA}
\affiliation{Fermi National Accelerator Laboratory, PO Box 500, Batavia, IL 60510, USA}

\author[0000-0002-5091-0470]{Samuel J. Schmidt}
\affiliation{University of California-Davis, Davis, CA 95616, USA}

\author[0000-0002-5631-0142]{Thomas D. Uram}
\affiliation{Argonne National Laboratory, Lemont, IL 60439, USA}

\author[0000-0002-8847-0335]{Antonio Villarreal}
\affiliation{Argonne National Laboratory, Lemont, IL 60439, USA}

\author[0000-0003-2035-2380]{Christopher W. Walter}
\affiliation{Department of Physics, Duke University, Durham NC 27708, USA }

\author[0000-0001-8653-7738]{Matthew P. Wiesner}
\affiliation{Benedictine University, Lisle, IL, 60532, USA}

\author[0000-0001-7113-1233]{W. Michael Wood-Vasey}
\affiliation{Department of Physics and Astronomy, University of Pittsburgh, Pittsburgh, PA 15260, USA}
\affiliation{Pittsburgh Particle Physics, Astrophysics and Cosmology Center (PITT PACC), University of Pittsburgh, Pittsburgh, PA 15260, USA}

\begin{abstract}

In preparation for cosmological analyses of the Vera C. Rubin Observatory Legacy Survey of Space and Time (LSST), the LSST Dark Energy Science Collaboration (LSST DESC) has created a 300~deg$^2$ simulated survey as part of an effort called Data Challenge 2 (DC2).  The DC2 simulated sky survey, in six optical bands with observations following a reference LSST observing cadence, was processed with the LSST Science Pipelines (19.0.0).  In this Note, we describe the public data release of the resulting object catalogs for the coadded images of five years of simulated observations along with associated truth catalogs. We include a brief description of the major features of the available data sets. To enable convenient access to the data products, we have developed a web portal connected to Globus data services. We describe how to access the data and provide example Jupyter Notebooks in Python to aid first interactions with the data.  We welcome feedback and questions about the data release via a GitHub repository.

\clearpage
\end{abstract}

\tableofcontents

\clearpage

\section{Introduction}

In the next decade, an unprecedented survey of the sky will be carried out using the Vera C. Rubin Observatory, the Legacy Survey of Space and Time \citep{2009arXiv0912.0201L,2019ApJ...873..111I}. One of the major aims of the survey is to unravel the origins of the accelerated expansion of the Universe. The LSST Dark Energy Science Collaboration (DESC)\footnote{\https{lsstdesc.org}} was formed to carry out this exciting endeavor~\citep{Abate:2012za}. In order to prepare for the arrival of data, LSST DESC has undertaken two data challenges (DC1 and DC2) based on sophisticated cosmological and  image simulations. The data challenges have been designed to mimic actual data from the Rubin Observatory in small, representative areas of the LSST observing footprint.

Both LSST DESC data challenges are based on realistic simulations of the extragalactic sky and employ an image simulation package, imSim, that provides access to a wide range of features. The resulting synthetic data were processed with Rubin's LSST Science Pipelines~\citep{2017ASPC..512..279J} to generate the final data products. The first data challenge, DC1, covers a $\sim$40 deg$^2$ area and ten years of observations. The image simulations were carried out for $r$-band only. A detailed description and a range of analysis results are provided in~\cite{dc1}.  In this data release note, we focus on the second data challenge, DC2. A comprehensive description of the LSST DESC DC2 Simulated Sky Survey can be found in~\cite{2020arXiv201005926L}. DC2 covers $\sim$300 deg$^2$ in the wide-fast-deep (WFD) area to be surveyed by LSST. Within this area, a small 1~deg$^2$ deep-drilling field (DDF), which has a much greater density of observations, has been simulated as well. For the data release described in this note, only data from the WFD campaign and 5 years of observations are provided, corresponding to the planned sixth Rubin data release, DR6. For DC2, all six optical bands $ugrizy$ are included in the image simulations. Both the extragalactic catalog and image simulations include many relevant features expected in LSST data at varying levels of realism, from simpler approximations to more realistic physical models.  The choice of how to represent the features depended on the complexity of the actual data and the finite resource availability. The DC2 overview paper \citep{2020arXiv201005926L} provides a comprehensive discussion about the DC2 design choices, which were guided mostly by considerations regarding cosmological probes.  

For LSST DESC, the data challenges serve multiple purposes. The advantage of simulated data is that the underlying truth is known. Therefore, even if they are not as complex as observational data or have different systematics, they provide an excellent testbed for DESC analysis pipelines. Given that the data formats closely mimic what is planned for the Rubin data products, they also serve to aid the development and optimization of data access methods. The data challenges also offer the opportunity to exercise the LSST Science Pipelines and investigate their performance, in particular with regard to how systematic effects in the data are handled. By making a first set of the data products publicly available, we hope that other LSST Science Collaborations will be able to carry out useful tests in preparation for arrival of LSST data as well. In addition, the data should be of value for the broader optical astronomy and cosmology communities.

This note is organized as follows. In~\autoref{sec:features} we describe the major features of the DC2 data set. We provide an overview of the data products that are part of this release in~\autoref{sec:products}. We provide instructions for data access, including a set of example Python Jupyter notebooks, in~\autoref{sec:access}. We conclude in~\autoref{sec:outlook} and provide a brief description of possible future data releases. 

\section{Major Features of the Data Set}
\label{sec:features}

\subsection{Astrophysical Inputs}
\label{sec:features:astro}
Here we describe the astrophysical inputs for the simulated WFD data set.\footnote{As described in \cite{2020arXiv201005926L}, data that were generated for the DDF region have additional astrophysical components such as strong lenses and an enhanced rate of transients.}  The components of this data set are mostly limited to the types of objects needed to support static probes of dark energy science, specifically the galaxies from the cosmoDC2 extragalactic catalog \citep{korytov}.  In addition, this data set includes stars from a simulated Milky Way, which are needed for astrometric and photometric calibration by the image processing pipeline, as well as Type Ia supernovae (SNe), which were included throughout the 300~deg$^2$ DC2 region.  The center of the WFD region is at R.A.~$= 61^\circ\!\!.863$, Decl.~$= -35^\circ\!\!.790$ (see~\autoref{fig:skymap}), and so the entire simulation region lies well outside of the Ecliptic and Galactic planes.
The coordinates of the corners of the WFD region can be found in Table~3 of \citet{2020arXiv201005926L}.

As noted, the galaxies are from the cosmoDC2 extragalactic catalog,\footnote{ \https{portal.nersc.gov/project/lsst/cosmoDC2/}} which covers 440 deg$^2$ out to a redshift of $z = 3$ and is complete to $m_r <28$.  The cosmoDC2 catalog is based on the Outer Rim $N$-body simulation \citep{2019ApJS..245...16H}, and the properties of the galaxies were derived using the Galacticus semi-analytic model \citep{benson_2010b} and painted onto dark matter halos using GalSampler \citep{2020MNRAS.495.5040H}.  The derived galaxy properties include stellar mass, morphology, spectral energy distributions, broadband filter magnitudes, host halo information, and weak lensing shears.   The bulge and disk components of the galaxies are rendered separately as S\'ersic profiles, and galaxies with $m_i < 27$ have ``knots'' of star formation added in order to model more complex light profiles for the brighter galaxies. The fluxes for these star-forming regions have been re-allocated from the disk component, and the knots have the same spectral energy distribution (SED) as the disk.

The Milky Way stars are simulated using the Galfast model of \citet{2008ApJ...673..864J}, which is based on densities and colors of stars in the Sloan Digital Sky Survey (SDSS). Stellar variability is included for periodic objects (e.g., RR Lyrae and Cepheids) and for non-periodic variables (e.g., CVs, flaring M-dwarfs, etc.).  Stars without a definitive variability class are modeled based on the Kepler Q17 data release \citep{2016ksci.rept....3T}.  The Galactic reddening is based on the three-dimensional model of \cite{2005AJ....130..659A}.

Finally, Type Ia SNe have been added throughout the DC2 region out to a redshift of $z=1.4$ with a population density that is consistent with observations~(e.g., \citealt{2010ApJ...713.1026D}). The simulated Type Ia SNe light curves were derived from a slightly modified version of the SALT2 model~\citep{2007A&A...466...11G}.

\subsection{Image Simulation Features}

DC2 used the \code{minion_1016} observing cadence,\footnote{ \https{docushare.lsst.org/docushare/dsweb/View/Collection-4604}} which was the Rubin Observatory LSST baseline cadence when production of the DC2 simulations began. This cadence provides the nominal field positions, telescope rotations, and filter selections for each 30-second pointing, as well as predicted seeing and airmass.  For the sky background, the LSST sky model is used, which is the ESO sky model with a twilight component added~\citep{2016SPIE.9910E..1AY}.  We have added random translational and rotational dithering to the nominal pointings to make the sky coverage more uniform.

The imSim simulation software, which is described in more detail in~\cite{2020arXiv201005926L}, uses the GalSim package \citep{2015A&C....10..121R} to render the astrophysical objects and the night sky.  The point-spread functions (PSFs) for each exposure are computed using a set of atmospheric phase screens that are realizations of Gaussian random fields with a Von Karman power spectrum.  In addition, the PSF calculation includes an optical model of the telescope based on modeling of the active optics system for Rubin Observatory.  After convolution with the PSF, objects are rendered on the LSST CCDs taking into account the instrumental throughput in each band, the object's SED within the bandpass, atmospheric effects such as differential chromatic refraction, and the convergence of the incident beam from the telescope optics. GalSim's sensor model includes the brighter-fatter and tree-ring electrostatic effects that are present in the CCDs used in the LSST Camera (LSSTCam).  Finally, electronics readout effects such as bleed trails, CCD segmentation, intra-CCD cross-talk, read noise, etc., are applied.  These effects are based on measurements of the actual LSSTCam hardware.

\subsection{Image Processing}
\label{sec:processing}
For the image processing of the DC2 data, we used version 19.0.0 of the LSST Science Pipelines code.\footnote{\https{pipelines.lsst.io/v/v19_0_0/index.html}}  The image processing steps are described in detail in \cite{2020arXiv201005926L}, \cite{10.1093/pasj/psx080}, and \cite{2018arXiv181203248B}.   Since the simulations lack throughput variation over the focal plane and from visit to visit, we omitted the joint photometric and astrometric calibration across visits, and consequently, the standard passbands for the LSST filters are simply the total throughputs\footnote{\https{github.com/lsst/throughputs/releases/tag/1.4}} used in the simulations and were derived by the Rubin systems engineering team.

\subsection{Known Issues}

\subsubsection{Ellipticity Distribution}

An issue with the intrinsic ellipticity distribution in the cosmoDC2 catalog has been identified and documented in \cite{Kovacs:2022:6336066}. This issue manifests in the distribution of intrinsic (not lensed) ellipticities being skewed to lower values, i.e., galaxies are rounder than they should have been. 
Note that the lensing distortions were unaffected, but they were added onto the too-round intrinsic shapes. Since the DC2 image simulation takes the lensed ellipticities in cosmoDC2 as input, the distribution of ellipticities in the DC2 object catalog (after image simulations) is also skewed to lower values, beyond the effect of the PSF.

\section{Available Data Sets}
\label{sec:products}

This Data Release (\thisversion{}) provides the data from the WFD campaign for 5 years of observations (corresponding to Rubin's DR6). This data set includes two tables: the Object Table (about 114 million extended sources and 33 million point sources; 118 gigabytes) and the Truth-match Table (about 759 million galaxy entries, 5 million star entries, and half million SN entries; 63 gigabytes). Each is partitioned by sky region (``tract'') into 166 files, as described in detail in ~\autoref{sec:representation}. In addition, two tracts of coadded images are also available. We define each of these data products in the subsections below. 

\subsection{Object Table}
\label{sec:object}

The Object Table contains information about static astronomical objects measured on a coadded image. The photometry in the Object Table is measured with the forced photometry method, i.e., it is consistently measured across multiple bands using a fixed position, which is determined from the reference band for each source \citep[Section~3.4 of][]{10.1093/pasj/psx080}. 

The generation of the Object Table is described in detail in Section~8 of \cite{2020arXiv201005926L}. In short, after the LSST Science Pipelines (19.0.0) produce the deepCoadd catalogs of multiple bands, we merge these catalogs across bands, and rename and compute certain columns to produce the final Object Table. The columns we rename or compute are meant to produce a science-ready catalog that resembles the Rubin LSST Data Products Definition Document (LSE-163; \https{lse-163.lsst.io}) for the end users. 

Each entry (row) in the Object Table corresponds to one measured astronomical object, assigned with a unique ID (\code{objectId}). There are no duplicated entries in the Object Table. For details about the file format and the organization of files for the Object Table, see \autoref{sec:representation-tables}. The full schema for the Object Table can be found in \autoref{app:object-schema}. 

\subsection{Truth-match Table}
\label{sec:truth}

The Truth-match Table is a joint representation of both the truth information (i.e., the perfect, noiseless measurement of astronomical objects' positions and fluxes within the standard passbands, used as inputs to the image simulations) and their best matches to the measured objects in the Object Table. The Truth-match Table allows users to examine, for example, the differences between true and measured fluxes and positions, and compare them to the expected levels of photometric and astrometric accuracy and precision.

The generation of the Truth-match Table is described in Section~4.2.1 of \cite{2020arXiv201005926L}. The truth information in the Truth-match Table only includes ``summary'' properties (i.e., static, or infinite-time averaged fluxes) of galaxies, stars, and SNe. Time-varying properties are not included. The match information stored in the  Truth-match Table is produced with the following procedure applied for each entry in the Object Table:
\begin{enumerate}
    \item Search for all truth entries that are within 1~arcsec and have an $r$-band magnitude difference ($\Delta r$) less than 1~mag. If one or more truth entries satisfying these criteria are found, pick the truth entry with the smallest $|\Delta r|$ as the match, and set \code{is_good_match} to \code{True}.
    \item If no truth entry was found in Step (1), pick the truth entry that is the nearest neighbor of the object entry on sky as the match, and set \code{is_good_match} to \code{False}.
\end{enumerate}
Given this procedure, every entry in the Object Table is assigned exactly one match. The majority of object entries have a ``good'' match (i.e., satisfying criteria in Step 1 above), and the rest have a nearest-neighbor match. More than 90\% of the ``good'' matches are not only the closest match in magnitude, but also the nearest neighbor match. 

As we will discuss further in \autoref{sec:representation}, the data tables are split by sky regions when stored on disk. The matching procedure described above was applied to each sky region individually; hence, a very tiny fraction ($<0.002\%$) of objects may not have good matches due to being too close to the sky region boundaries. 

Because the Object Table was used as the reference catalog for the matching procedure, some truth entries may be chosen as a match more than once, while others may not be chosen at all.
Flags to distinguish these situations are included in the Truth-match Table. 
Selecting all entries with \code{match_objectId} $> -1$ from the Truth-match Table would result in a subset of Truth-match entries that have exactly the same row order as the entries in the Object Table (and hence may contain duplicated truth entries). On the other hand, selecting all entries with \code{is_unique_truth_entry} being \code{True} would produce a subset of Truth-match entries that contains all unique truth entries, including truth entries that have not been chosen as a match.
These selections are particularly important for users who wish to access the files directly. For users who use GCRCatalogs (\autoref{sec:gcr}), the reader will automatically select the correct rows depending on whether the user loads the ``Truth'' view (having all truth entries) or the ``Match'' view (having entries that matches to the Object Table). We refer users to the example notebooks we provided (\autoref{sec:notebooks}) for detail.

The file format and the organization of files for the Truth-match Table are described in \autoref{sec:representation-tables}. The full schema for the Truth-match Table can be found in \autoref{app:truth-schema}.

\subsection{Unmerged Truth Information}
\label{sec:truth-unmerged}

Characteristics of the three types of sources (galaxies, stars, and SNe) can be found in several unmerged tables. These tables provide additional information about these sources beyond what is included in the Truth-match table. Users can use truth object IDs to match between these tables. 

The static truth information of galaxies is stored in a single table. For the variable sources (stars and SNe), the truth information is stored in two types of tables:
\begin{enumerate}
    \item Summary Tables, one each for stars and SNe. In these tables, each row represent one single object (star or SN). These tables are similar to the Truth-match table but
    \begin{enumerate}
        \item they are restricted to sources which are stars or SNe, respectively
        \item they do not include columns which have to do with matching or spatial partitioning
        \item each has a few extra columns specific to the source type
        \item Truth-match columns whose value is constant for a particular source type (e.g., \code{is_variable} for SNe) are omitted in the unmerged file
    \end{enumerate}
    \item Variability Tables, one each for stars and SNe. These tables have a row for each observation of a variable source, excluding stars whose variability would not be detectable by LSST. Note that most stars are classed as ``variable'' but only about 10\% vary enough to be included in the variability table; for such stars, the flag field \code{above_threshold} in the star summary table is set. These tables have only a few columns, including ID, which may be matched with the corresponding Summary Table ID.
   
\end{enumerate}

For details about file format and organization see \autoref{sec:representation-unmerged}. The complete schema for these tables is described Tables~\ref{app:star-summary-truth-schema}--\ref{app:sn-variability-truth-schema}.

\subsection{Intermediate Data Products from LSST Science Pipelines}
\label{sec:intermediate-data-products}

We provide the calibrated exposures, background-subtracted images, and source catalogs produced by the single-epoch processing as well as the coadded images (and the corresponding calibrated exposures, background models, and detected sources; see \autoref{app:intermediate-data-products}) for a small sky region (tracts 3828 and 3829; see \autoref{sec:representation} for definition). These images are part of the direct outputs from the image processing using the LSST Science Pipelines (see \autoref{sec:processing}). The single-epoch and coaddition processes are described in Sec.~3.1 and Sec.~3.3 of \cite{10.1093/pasj/psx080}, respectively, as well as in Sec.~3.1 and Sec.~4.2 of \cite{2018arXiv181203248B}. 
The file format and the organization of files for the single-epoch and coadded images are also described in \autoref{sec:representation-intermediate} and \autoref{app:intermediate-data-products}.

\section{Data Access}
\label{sec:access}

\subsection{Data File Format and Organization}
\label{sec:representation}

\subsubsection{Object and Truth-match Tables}
\label{sec:representation-tables}

All data tables in this release are stored in the Apache Parquet format,\footnote{\https{parquet.apache.org}} an efficient columnar storage form, with I/O tools readily available for multiple development systems. 
The data files can be easily downloaded to the user's machine via Globus (\autoref{sec:download}), and read with Python packages such as \code{pyarrow} or \code{fastparquet}.\footnote{\code{pyarrow}: \https{arrow.apache.org/docs/python}; \code{fastparquet}: \https{fastparquet.readthedocs.io}}
We additionally provide a Python package, \code{GCRCatalogs}, which contains a high-level user interface to access the data files (see \autoref{sec:gcr}).

\begin{figure}[tbh!]
    \centering
    \includegraphics[width=0.8\textwidth]{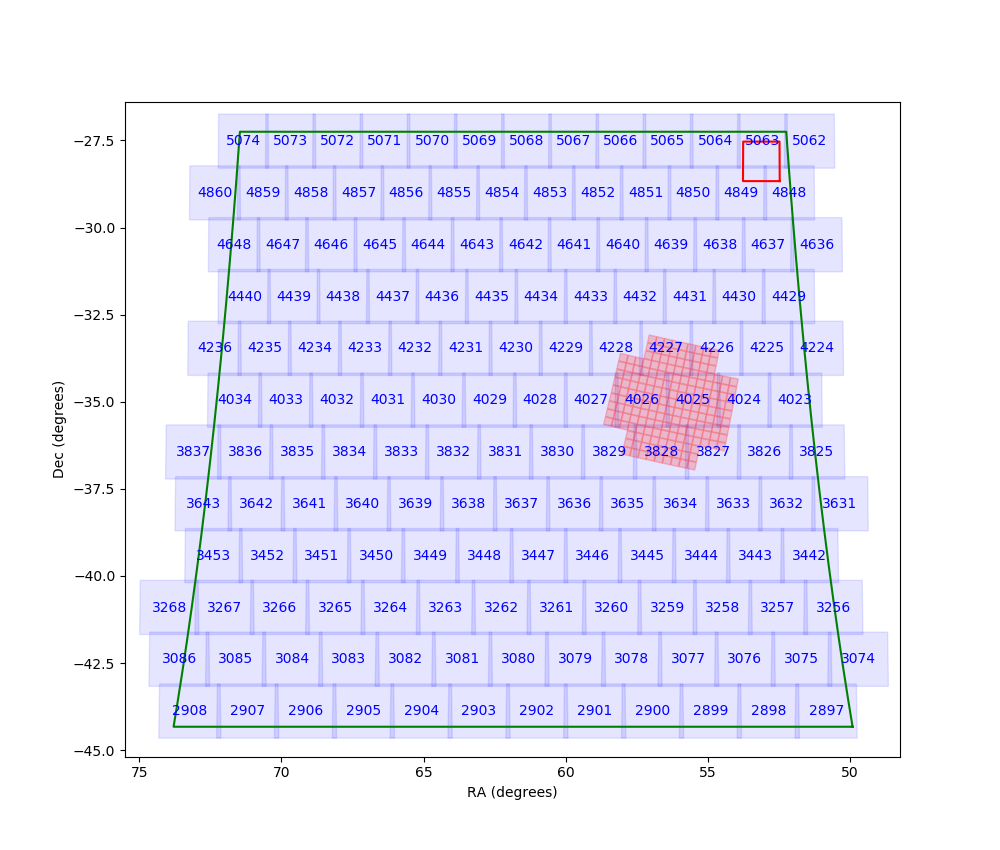}
    \caption{Sky map of the DC2 footprint. The large green trapezoid is the full DC2 footprint. The small red square in the upper right corner is the DDF region that is excluded in this release. Each tract is represented by a rectangle with a number on it showing the tract ID. The light pink region shows the size of the LSSTCam focal plane as a comparison.}
    \label{fig:skymap}
\end{figure}

Each data table is further partitioned into several files that correspond to different parts of the sky. The partition is based on the ``tract'' value in the ``Rings sky map'' pixelization of LSST Science Pipelines.\footnote{\https{pipelines.lsst.io/py-api/lsst.skymap.ringsSkyMap.RingsSkyMap.html}} \autoref{fig:skymap} illustrates the partition. The same partition is used for both Object Table and Truth-match Table. No padding is included; i.e., an entry that is near a tract boundary still only appears in the tract it belongs to. Each Parquet file contains only one partition (row group).

\subsubsection{Unmerged Truth Tables}
\label{sec:representation-unmerged}

Each of the four point-source tables (star summary, star variability, SN summary, SN variability) is available in two formats: SQLite and Apache Parquet.  The table schema is identical between the two formats.  These files are not partitioned by sky area. 

The point-source Parquet files are smaller than the SQlite versions; particularly in the case of the variability truth tables (SQlite versions are about 10 times larger).  Also the Parquet files, and only the Parquet files, may be accessed via the Python package \code{GCRCatalogs}. However, some applications, such as querying light curves of specific objects, will greatly benefit from the efficiency provided by the SQLite files, which are indexed by object and visit IDs.

The galaxy summary table is available only in Parquet, partitioned into healpix pixels.  

\subsubsection{Intermediate Data Products from LSST Science Pipelines}
\label{sec:representation-intermediate}

The intermediate data products we provide in this release are parts of the output files from v19.0.0 of the LSST Science Pipelines and they are stored in the original formats from the LSST Science Pipelines. Images are stored in the FITS (Flexible Image Transport System)format,\footnote{\https{fits.gsfc.nasa.gov/fits_standard.html}} and partitioned by filter (passband) and sky region. The sky region partition scheme follows the same ``Rings sky map'' described above. Each tract is further subdivided into ``patches'', and each individual coadded image file corresponds to a single patch. The descriptions of files included in this data set, and the associated filename patterns, can be found in \autoref{app:intermediate-data-products}.

\subsection{Downloading Data Files}
\label{sec:download}

The data files are made available via a data portal website\footnote{\https{lsstdesc-portal.nersc.gov}\label{fn:portal}} using Globus\footnote{\https{www.globus.org}} services. Any user can authenticate using their organizational login or with a Globus ID\footnote{\https{www.globusid.org/what}} to initiate transfers of the full or partial data set to another endpoint to which the user has access, whether it be a laptop or a high-performance computing center. 

Detailed instructions on data transfers can be found on our data portal website (see Footnote~\ref{fn:portal}).

\subsection{Accessing Data in Python}
\label{sec:gcr}

While the tabular data files are accessible via standard Parquet I/O tools, we provide a high-level Python package, \code{GCRCatalogs},\footnote{\https{github.com/LSSTDESC/gcr-catalogs}} to assist users to access DC2 data. 

The \code{GCRCatalogs} package is installable by package managers \code{pip} and \code{conda}.
Once installed, the package should be configured to recognize the path to which the data files have been downloaded. The DC2 data set will then be readily available in the user's own Python environment. 
Detailed instructions can be found in our data portal website (see Footnote~\ref{fn:portal}).

Below we show an example Python code snippet to demonstrate how to use \code{GCRCatalogs} to obtain the R.A. and Decl. columns from three tracts of the Object Table, with a sampling rate of 1\%. The high-level user interface makes the code simple and readable. 

\begin{minted}[frame=lines,samepage]{python}
import GCRCatalogs
from GCRCatalogs.helpers.tract_catalogs import tract_filter, sample_filter

obj_cat = GCRCatalogs.load_catalog("desc_dc2_run2.2i_dr6_object")

data = obj_cat.get_quantities(
    quantities=['ra', 'dec'],       # columns we want to load, 
    filters=[sample_filter(0.01)],  # down sample at 1%
    native_filters=[tract_filter([4225, 4226, 4430])],  # select 3 tracts
)
\end{minted}

For accessing the coadded images, we recommend using standard FITS I/O tools, such as the \code{fits} module in \code{astropy}.\footnote{\https{docs.astropy.org/en/stable/io/fits/index.html}}

\subsection{Example Python Jupyter Notebook}
\label{sec:notebooks}

\begin{figure}[tbh!]
    \centering
    \includegraphics[width=\textwidth]{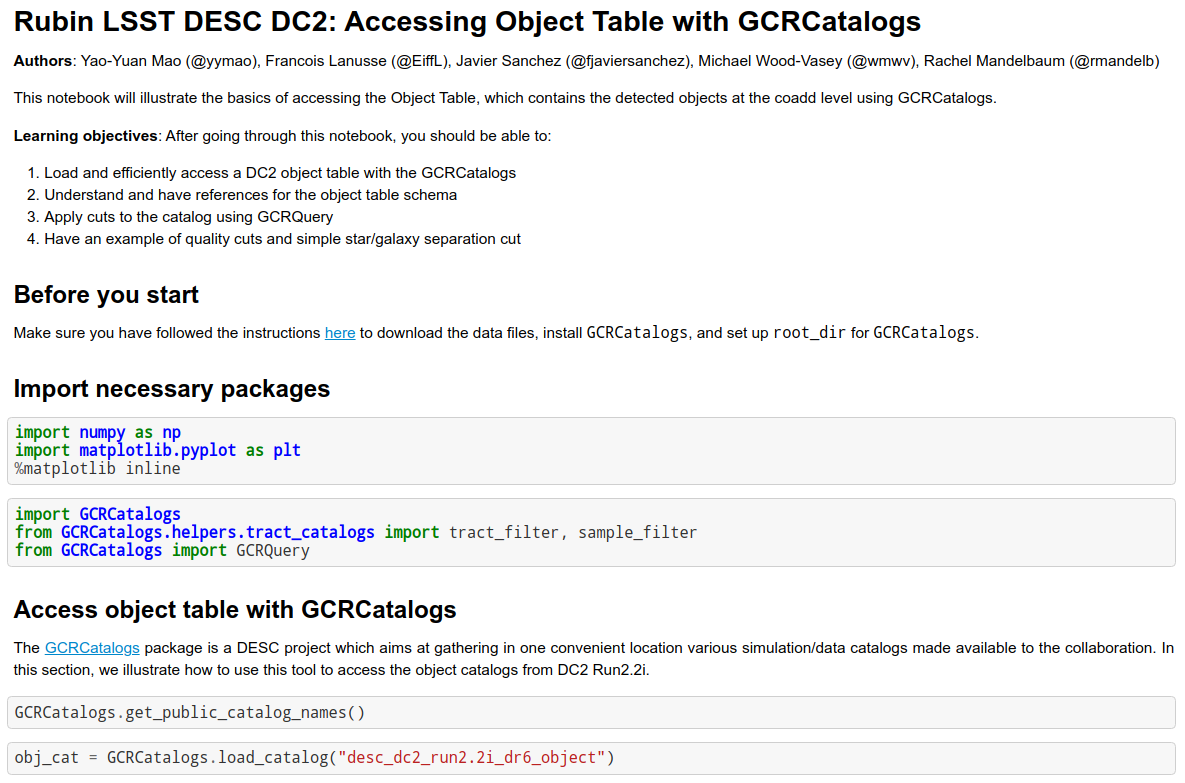}
    \caption{Screenshot of the beginning of the Object Table tutorial notebook.}
    \label{fig:notebook}
\end{figure}
\begin{figure}[tbh!]
    \centering
    \includegraphics[width=0.75\textwidth]{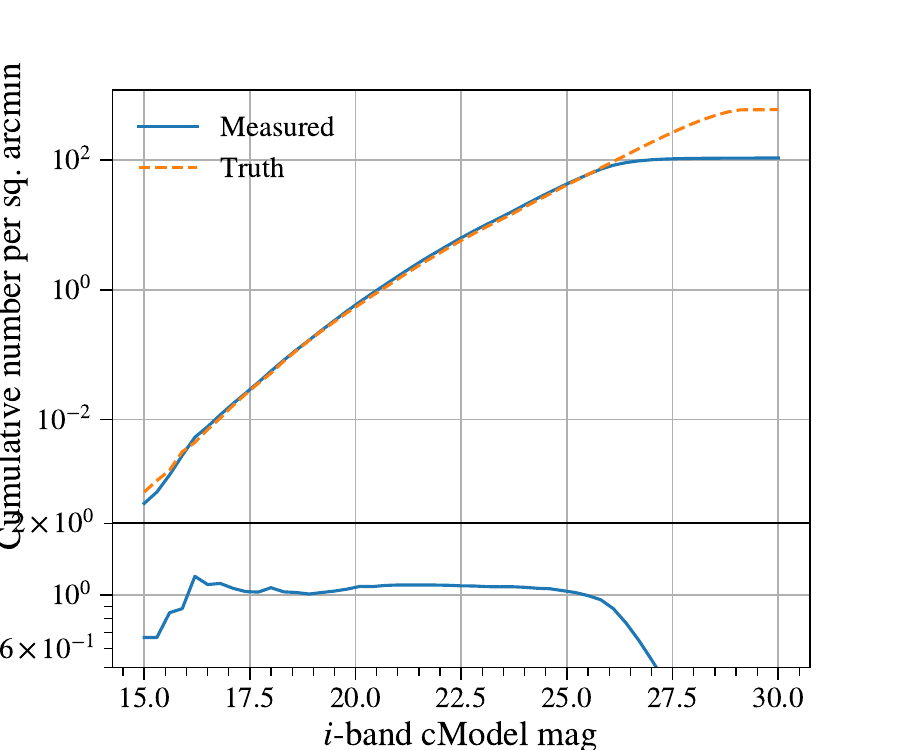}
    \caption{A resulting plot from the example analysis in the tutorial notebook for the Truth-match Table. The plot shows the measured and true galaxy number density as a function of magnitude. The code to generate this plot is included in the notebook.}
    \label{fig:number_density}
\end{figure}

We provide two example Python Jupyter Notebooks which further demonstrate how to use \code{GCRCatalogs} to access the tabular data 
(currently no example notebook is available for coadded images).
They are designed as tutorials, with clear instructions (see a screenshot in \autoref{fig:notebook}). In addition to demonstrating the access method, these notebooks also explain the data model in detail, and provide a few simple analyses that can be used as starting points for further development (see an example plot from the notebook in \autoref{fig:number_density}).
These notebooks can be found in a GitHub repository.\footnote{\https{github.com/LSSTDESC/desc-data-portal/tree/main/notebooks}}

\section{Summary and Outlook}
\label{sec:outlook}
In this Note we described the first public data release for DC2 carried out by the LSST DESC. We make data available for a simulated WFD survey spanning 300 degree$^2$ and 5 years of Rubin observations, including a subset of the coadd-based catalogs that would be included in Rubin Observatory's DR6. 

This data release (\thisversion{}) focuses on a limited set of data products generated with the LSST Science Pipelines. In the future, we plan to extend this data release in several directions. First, LSST DESC is currently working on generating so-called ``add-on'' catalogs. These catalogs provide information obtained from further processing the data. Examples include a photo-$z$ catalog and a cluster catalog. Once these catalogs have been carefully validated and are of sufficient quality to be of broader interest, they will be added to the DC2 Data Release. Second, the processing of the DDF portion of DC2 is still in progress. As explained in more detail in~\cite{2020arXiv201005926L}, the DDF region contains several astrophysical components, e.g.~AGNs, that are not available in the WFD region. As with the add-on catalogs, once careful validation has concluded, we plan to make those data available as well. Finally, for cosmology it is very informative to compare results from different survey data releases to build a better understanding of the impact of the depth of the data on cosmological constraints. Therefore, LSST DESC is currently generating additional coadds and associated catalogs for subsets of the data corresponding to 1- and 2-year depths. Depending on the feedback we receive, these datasets may become part of future public data releases as well.

\section{Conditions of Use}

If you use these data products in a publication, we ask that you cite this Note and the DC2 overview paper~\citep{2020arXiv201005926L}.
If you use \code{GRCCatalogs} to access the data products, please also cite \cite{2018ApJS..234...36M}. 
If you would like to serve these data sets on a mirror site, or to ingest and serve them via a different access method (e.g., a database), we ask you to inform LSST~DESC at \href{mailto:papers@lsstdesc.org}{papers@lsstdesc.org} so that we can contact you when future updates are made to these data sets.
We encourage all users to provide feedback and ask questions via the ``Discussions'' feature in the dedicated GitHub repository.\footnote{\https{github.com/LSSTDESC/desc-data-portal/discussions}}

\clearpage
\appendix
\section{Changelog}
\begin{ThreePartTable}
\begin{TableNotes}
\footnotesize
\item[] ~
\end{TableNotes}
\begin{longtable}{p{0.8in}p{0.8in}p{4in}}
\hline
\textbf{Date} & \textbf{Version} & \textbf{Description} \\ 
\hline
\endhead
\endfoot
\hline
\insertTableNotes  
\endlastfoot
06/16/2025 & v5 & The center coordinates of the WFD region stated in \autoref{sec:features:astro} are corrected. \\
06/15/2022 & v4 & (1) New data sets available in this release: (a) unmerged truth catalogs (summaries for SNe, stars and galaxies and variability information for stars and SNe; \autoref{sec:truth-unmerged}), and (b) additional intermediate data products (calibrated exposures, background subtracted images, and source catalogs output from the single-epoch processing;  \autoref{sec:intermediate-data-products}). (2) Truth table and Truth-match table have been updated to match the unmerged truth catalogs. (3) ``Known Issues'' section was added to this note. \\
12/10/2021 & v3 & The URL to DESC Data Portal was updated. \\
06/08/2021 & v2 & A minor processing issue was fixed in Tract 4852; the object and truth-match tables are updated. Two tracts of coadded images are provided. \\
01/07/2021 & v1 & Initial release \\
\end{longtable}
\end{ThreePartTable}

\clearpage

\section{Table Schema}

\subsection{Object Table Schema}
\label{app:object-schema}
\begin{ThreePartTable}
\begin{TableNotes}
\footnotesize
\item [\hypertarget{obj_fn1}{1}] In LSE-163, \code{I<xx,yy,xy>} and \code{I<xx,yy,xy>PSF} are defined in the units of squared arcsec. 
\end{TableNotes}
\begin{longtable}{p{1.7in}p{0.5in}p{0.6in}p{2.8in}}
\hline
\textbf{Name} & \textbf{Type} & \textbf{Unit} & \textbf{Description} \\ 
\hline
\endhead
\endfoot
\hline
\insertTableNotes  
\endlastfoot
\code{objectId} & int64 & -- & Unique object ID \\
\code{parentObjectId} & int64 & -- & Parent object ID \\
\code{good} & bool & -- & \code{True} if the source has no flagged pixels \\
\code{clean} & bool & -- &  \code{True} if the source has no flagged pixels (i.e., \code{good}) and is not skipped by the deblender \\
\code{blendedness} & float64 & -- & Measure of how flux is affected by neighbors ($1 - I_\text{child}/I_\text{parent}$; see \S\,4.9.11 of \citealt{10.1093/pasj/psx080}) \\
\code{extendedness} & float64 & -- & 0 for stars; 1 for extended objects \\
\code{ra} & float64 & degree & Right Ascension \\
\code{dec} & float64 & degree & Declination \\
\code{x} & float64 & pixel & 2D centroid location (x coordinate) \\
\code{y} & float64 & pixel & 2D centroid location (y coordinate) \\
\code{xErr} & float32 & pixel & Error value for \code{x} \\
\code{yErr} & float32 & pixel & Error value for \code{y} \\
\code{xy_flag} & bool & -- & Flag for issues with \code{x} and \code{y} \\
\code{tract} & int64 & -- & Tract ID in Sky Map \\ 
\code{patch} & string & -- & Patch ID in Sky Map (as a string, \code{`x,y'})\\ 
\code{Ixx_pixel} & float64 & sq.~pixel$^\text{\hyperlink{obj_fn1}{1}}$ & Adaptive second moment ($xx$) of source intensity, averaged across bands \\
\code{Ixx_pixel_<band>} & float64 & sq.~pixel$^\text{\hyperlink{obj_fn1}{1}}$ & Adaptive second moment ($xx$) of source intensity in \code{<band>} \\
\code{Iyy_pixel} & float64 & sq.~pixel$^\text{\hyperlink{obj_fn1}{1}}$ & Adaptive second moment ($yy$) of source intensity, averaged across bands \\
\code{Iyy_pixel_<band>} & float64 & sq.~pixel$^\text{\hyperlink{obj_fn1}{1}}$ & Adaptive second moment ($yy$) of source intensity in \code{<band>} \\
\code{Ixy_pixel} & float64 & sq.~pixel$^\text{\hyperlink{obj_fn1}{1}}$ & Adaptive second moment ($xy$) of source intensity, averaged across bands \\
\code{Ixy_pixel_<band>} & float64 & sq.~pixel$^\text{\hyperlink{obj_fn1}{1}}$ & Adaptive second moment ($xy$) of source intensity in \code{<band>} \\
\code{I_flag} & bool & -- & Flag for issues with \code{Ixx}, \code{Iyy_pixel}, and \code{Ixy} \\
\code{I_flag_<band>} & bool & -- & Flag for issues with \code{Iyy_pixel_<band>}, \code{Ixy_<band>}, and \code{Ixx_<band>} \\
\code{IxxPSF_pixel} & float64 & sq.~pixel$^\text{\hyperlink{obj_fn1}{1}}$ & Adaptive second moment ($xx$) of PSFy, averaged across bands \\
\code{IxxPSF_pixel_<band>} & float64 & sq.~pixel$^\text{\hyperlink{obj_fn1}{1}}$ & Adaptive second moment ($xx$) of PSF in \code{<band>} \\
\code{IyyPSF_pixel} & float64 & sq.~pixel$^\text{\hyperlink{obj_fn1}{1}}$ & Adaptive second moment ($yy$) of PSFy, averaged across bands \\
\code{IyyPSF_pixel_<band>} & float64 & sq.~pixel$^\text{\hyperlink{obj_fn1}{1}}$ & Adaptive second moment ($yy$) of PSF in \code{<band>} \\
\code{IxyPSF_pixel} & float64 & sq.~pixel$^\text{\hyperlink{obj_fn1}{1}}$ & Adaptive second moment ($xy$) of PSFy, averaged across bands \\
\code{IxyPSF_pixel_<band>} & float64 & sq.~pixel$^\text{\hyperlink{obj_fn1}{1}}$ & Adaptive second moment ($xy$) of PSF in \code{<band>} \\
\code{psf_fwhm_<band>} & float64 & arcsec & PSF FWHM calculated from \code{base_SdssShape} \\
\code{psNdata} & float32 & - & Number of data points (pixels)
used to fit the model \\
\code{psFlux_<band>} & float64 & nJy & Point-source model flux in \code{<band>} \\
\code{psFluxErr_<band>} & float64 & nJy & Error value for \code{psFlux_<band>} \\
\code{psFlux_flag_<band>} & bool & -- & Flag for issues with \code{psFlux_<band>} \\
\code{mag_<band>} & float64 & AB mag & Point-source model magnitude in \code{<band>}\\
\code{magerr_<band>} & float64 & AB mag & Error value for \code{mag_<band>} \\
\code{cModelFlux_<band>} & float64 & nJy & Composite model (cModel) flux in \code{<band>} \\
\code{cModelFluxErr_<band>} & float64 & nJy & Error value for \code{cModelFlux_<band>} \\
\code{cModelFlux_flag_<band>} & bool & -- & Flag for issues with \code{cModelFlux_<band>} \\
\code{mag_<band>_cModel} & float64 & AB mag & cModel magnitude in \code{<band>} \\
\code{magerr_<band>_cModel} & float64 & AB mag & Error value for \code{mag_<band>_cModel} \\
\code{snr_<band>_cModel} & float64 & -- & Signal-to-noise ratio for cModel magnitude in \code{<band>} \\
\end{longtable}
\end{ThreePartTable}

\bigskip

\subsection{Truth-match Table Schema}
\label{app:truth-schema}
\begin{ThreePartTable}
\begin{TableNotes}
\footnotesize
\item [\hypertarget{truth_fn1}{1}] When accessing this catalog as \code{dc2_object_run2.2i_dr6_wfd_with_truth_match} via \code{GCRCatalogs}, all the columns in this table, except for the last six, are postfixed with \code{_truth}.
\item [\hypertarget{truth_fn2}{2}] Only present when accessing the catalog via \code{GCRCatalogs}.
\item [\hypertarget{truth_fn3}{3}] Because the object catalog is used as the reference catalog for matching, some truth entries may appear more than once, and some truth entries may not have a matching object.
\end{TableNotes}
\begin{longtable}{p{1.6in}p{0.5in}p{0.6in}p{2.9in}}
\hline
\textbf{Name}$^\text{\hyperlink{truth_fn1}{1}}$ & \textbf{Type} & \textbf{Unit} & \textbf{Description} \\ 
\hline
\endhead
\endfoot
\hline
\insertTableNotes  
\endlastfoot
\code{id} & int64 & -- & Unique object ID \\ 
\code{id_string} & string & -- & Unique object ID in string type \\ 
\code{host_galaxy} & int64 & -- & ID of the host galaxy for a SN entry ($-1$ for other truth types)\\ 
\code{truth_type} & int32 & -- & 1 for galaxies, 2 for stars, and 3 for SNe \\ 
\code{ra} & float64 & degree & Right Ascension \\
\code{dec} & float64 & degree & Declination \\
\code{redshift} & float64 & -- & Redshift \\ 
\code{flux_<band>} & float32 & nJy & Static flux value in \code{<band>} \\ 
\code{mag_<band>}$^\text{\hyperlink{truth_fn2}{2}}$ & float32 & AB mag & Magnitude in \code{<band>} \\ 
\code{av} & float32 & -- & Milky Way total extinction parameter $A(V)$ at object location \\ 
\code{rv} & float32 & -- &  Milky Way relative visibility parameter $R(V) = A(V)/E(B-V)$ at object location \\ 
\code{tract} & int64 & -- & Tract ID in Sky Map \\ 
\code{patch} & string & -- & Patch ID in Sky Map (as a string, \code{`x,y'}) \\ 
\code{cosmodc2_hp} & int64 & -- & Healpix ID in cosmoDC2 (for galaxies only; $-1$ for stars and SNe) \\ 
\code{cosmodc2_id} & int64 & -- & Galaxy ID in cosmoDC2 (for galaxies only; $-1$ for stars and SNe)\\ 
\code{match_objectId} & int64 & -- & \code{objectId} of the matching object entry ($-1$ for unmatched truth entries$^\text{\hyperlink{truth_fn3}{3}}$) \\ 
\code{match_sep} & float64 & arcsec & On-sky angular separation of this object--truth matching pair ($-1$ for unmatched truth entries$^\text{\hyperlink{truth_fn3}{3}}$) \\ 
\code{is_good_match} & bool & -- & \code{True} if this object--truth matching pair satisfies all matching criteria \\
\code{is_nearest_neighbor} & bool & -- & \code{True} if this truth entry is the nearest neighbor of the object specified by \code{match_objectId} \\
\code{is_unique_truth_entry} & bool & -- & \code{True} for truth entries that appear for the first time in this truth table$^\text{\hyperlink{truth_fn3}{3}}$ \\
\end{longtable}
\end{ThreePartTable}

\bigskip

\subsection{Star Summary Truth Table Schema}
\label{app:star-summary-truth-schema}
\begin{ThreePartTable}
\begin{TableNotes}
\end{TableNotes}
\begin{longtable}{p{1.6in}p{0.5in}p{0.6in}p{2.9in}}
\hline
\textbf{Name} & \textbf{Type} & \textbf{Unit} & \textbf{Description} \\ 
\hline
\endhead
\code{id} & int64 & -- & Unique object ID\\
\code{ra} & float64 & degree & Right Ascension\\
\code{dec} & float64 & degree & Declination\\
\code{flux_<band>} & float32 & nJy & Static flux value in <band>\\
\code{model} & string & -- & variability model\\
\code{max_stdev_delta_mag} & float32 & AB mag & max stdev in delta magnitude across all bands\\
\code{above_threshold} & bool & -- & True if above quantity satisfies threshold (here set to 1 mmag) for LSST detectability\\
\code{av} & float32 & -- & Milky Way total extinction parameter $A(V)$ at object location \\ 
\code{rv} & float32 & -- &  Milky Way relative visibility parameter $R(V) = A(V)/E(B-V)$ at object location \\
\hline
\end{longtable}
\end{ThreePartTable}

\bigskip

\subsection{Supernova Summary Truth Table Schema}
\label{app:sn-summary-truth-schema}
\begin{ThreePartTable}
\begin{longtable}{p{1.6in}p{0.5in}p{0.6in}p{2.9in}}
\hline
\textbf{Name} & \textbf{Type} & \textbf{Unit} & \textbf{Description} \\ 
\hline
\endhead
\code{id_string} & string & -- & Unique object ID\\
\code{id} & int64 & -- & alternate id which is an int\\
\code{host_galaxy} & int64 & -- & associated galaxy\\
\code{ra} & float64 & degree & Right Ascension\\
\code{dec} & float64 & degree & Declination\\
\code{redshift} & float64 & -- & Redshift\\
\code{mB} & float64 & -- & normalization factor \\
\code{c} & float64 & -- & empirical parameter controlling the colors\\
\code{t0} & float64 & -- & time in mjd of phase=0, corresponding to B-band maximum \\
\code{x0} & float64 & -- & normalization factor\\
\code{x1} & float64 & -- & empirical parameter controlling the stretch in time of light curves\\
\code{max_flux_<band>} & float32 & nJy & max flux observed for <band>\\
\code{av} & float32 & -- & Milky Way total extinction parameter $A(V)$ at object location \\ 
\code{rv} & float32 & -- &  Milky Way relative visibility parameter $R(V) = A(V)/E(B-V)$ at object location \\
\hline
\end{longtable}
\end{ThreePartTable}

\bigskip

\subsection{Galaxy Summary Truth Table Schema}
\label{app:galaxy-summary-truth-schema}
\begin{ThreePartTable}
\begin{TableNotes}
\end{TableNotes}
\begin{longtable}{p{1.6in}p{0.5in}p{0.6in}p{2.9in}}
\hline
\textbf{Name} & \textbf{Type} & \textbf{Unit} & \textbf{Description} \\ 
\hline
\endhead
\code{id} & int64 & -- & Unique object ID\\
\code{host_galaxy} & int64 & -- & Always -1 for galaxies\\
\code{ra} & float64 & degree & Right Ascension\\
\code{dec} & float64 & degree & Declination\\
\code{redshift} & float64 & -- & Redshift\\
\code{is_variable} & bool & -- & Always False for galaxies\\
\code{is_pointsource} & bool & -- & Always False for galaxies\\

\code{flux_<band>} & float32 & nJy & Static flux value in <band>\\
\code{flux_<band>_noMW} & float32 & nJy & Static flux value in <band>, without Milky Way extinction (i.e., dereddened)\\
\code{av} & float32 & -- & Milky Way total extinction parameter $A(V)$ at object location \\ 
\code{rv} & float32 & -- &  Milky Way relative visibility parameter $R(V) = A(V)/E(B-V)$ at object location \\
\hline
\end{longtable}
\end{ThreePartTable}

\bigskip

\subsection{Star Variability Truth Table Schema}
\label{app:star-variability-truth-schema}
\begin{ThreePartTable}
\begin{TableNotes}
\end{TableNotes}
\begin{longtable}{p{1.6in}p{0.5in}p{0.6in}p{2.9in}}
\hline
\textbf{Name} & \textbf{Type} & \textbf{Unit} & \textbf{Description} \\ 
\hline
\endhead
\code{id} & int64 & -- & Unique object ID\\
\code{obsHist} & int64 & -- & unique visit ID\\
\code{MJD} & float64 & day & time of visit\\
\code{bandpass} & string & -- & LSST band\\
\code{delta_flux} & float32 & nJy & delta from \code{flux_<band>} in summary table\\
\end{longtable}
\end{ThreePartTable}

\bigskip

\subsection{Supernova Variability Truth Table Schema}
\label{app:sn-variability-truth-schema}
\begin{ThreePartTable}
\begin{longtable}{p{1.6in}p{0.5in}p{0.6in}p{2.9in}}
\hline
\textbf{Name} & \textbf{Type} & \textbf{Unit} & \textbf{Description} \\ 
\hline
\endhead
\code{id} & int64 & -- & Unique integer object ID, unrelated to \code{id_string}\\
\code{obsHist} & int64 & -- & unique visit ID\\
\code{MJD} & float64 & day & time of visit\\
\code{bandpass} & string & -- & LSST band\\
\code{delta_flux} & float32 & nJy & flux for this visit (``average'' is 0 for SNe)\\
\code{id_string} & string & -- & unique object ID\\
\end{longtable}
\end{ThreePartTable}

\bigskip
\section{Intermediate Data Products File Descriptions}
\label{app:intermediate-data-products}
\begin{ThreePartTable}
\begin{TableNotes} 
\item
\end{TableNotes}
\begin{longtable}{p{6.1in}}
\hline
\code{Filename} \code{pattern} \\
\quad \textit{Description} \\ 
\hline \hline
\endhead
\endfoot
\hline
\insertTableNotes  
\endlastfoot
\code{raw/{visit}/{raft}/{visit}-{raft}-{sensor}-{detector}.fits} \\
\quad \textit{Raw simulated visits} \\ \hline
\code{calexp/{visit}-{band}/{raft}/calexp_{visit}-{band}-{raft}-{sensor}-{detector}.fits} \\
\quad \textit{Calibrated exposures} \\ \hline
\code{calexp/{visit}-{band}/{raft}/bkgd_{visit}-{band}-{raft}-{sensor}-{detector}.fits} \\
\quad \textit{Background subtracted images} \\ \hline
\code{src/{visit}-{band}/{raft}/src_{visit}-{band}-{raft}-{sensor}-{detector}.fits} \\
\quad \textit{Single-epoch Source Catalogs} \\ \hline
\code{deepCoadd/{filter}/{tract}/{patch}.fits} \\
\quad \textit{Coadded images} \\ \hline
\code{deepCoadd/{filter}/{tract}/{patch}_nImage.fits} \\
\quad \textit{Number of visit-level images contributing to each pixel of the coadded image} \\ \hline
\code{deepCoadd-results/{filter}/{tract}/{patch}/bkgd-{filter}-{tract}-{patch}.fits} \\
\quad \textit{Background model for the calibrated exposure of the coadded image} \\ \hline
\code{deepCoadd-results/{filter}/{tract}/{patch}/calexp-{filter}-{tract}-{patch}.fits} \\
\quad \textit{Calibrated exposure of the coadded image} \\ \hline
\code{deepCoadd-results/{filter}/{tract}/{patch}/det-{filter}-{tract}-{patch}.fits} \\
\quad \textit{Source catalog of footprints detected on the coadded image} \\ \hline
\end{longtable}
\end{ThreePartTable}

\clearpage
\section*{Acknowledgments}
\phantomsection
\addcontentsline{toc}{section}{Acknowledgments}

The DESC acknowledges ongoing support from the Institut National de 
Physique Nucl\'eaire et de Physique des Particules in France; the 
Science \& Technology Facilities Council in the United Kingdom; and the
Department of Energy, the National Science Foundation, and the LSST 
Corporation in the United States.  DESC uses resources of the IN2P3 
Computing Center (CC-IN2P3--Lyon/Villeurbanne - France) funded by the 
Centre National de la Recherche Scientifique; the National Energy 
Research Scientific Computing Center, a DOE Office of Science User 
Facility supported by the Office of Science of the U.S.\ Department of
Energy under Contract No.\ DE-AC02-05CH11231; STFC DiRAC HPC Facilities, 
funded by UK BIS National E-infrastructure capital grants; and the UK 
particle physics grid, supported by the GridPP Collaboration.  This 
work was performed in part under DOE Contract DE-AC02-76SF00515.

The work of SH, APH, KH, JH, EK, DK, PL, TU and ASV at Argonne National Laboratory was supported under the U.S. DOE contract DE-AC02-06CH11357.
Support for YYM was provided by NASA through the NASA Hubble Fellowship grant no.\ HST-HF2-51441.001 awarded by the Space Telescope Science Institute, which is operated by the Association of Universities for Research in Astronomy, Incorporated, under NASA contract NAS5-26555. 


BA investigated variations in the sky model across the focal plane in imSim.
HA implemented the dithers and extracted the visit lists for the simulations.
YNB worked on the design and implementation of the imSim workflow and developed extensions to Parsl to meet the performance and scalability needs of the imSim workflow.
FEB contributed to the development and testing of the AGN model.
GB managed the European computational grid work for DC2.
RB conceptualized the interaction of Time Domain Science implementations with existing middleware software, compiled scientific desiderata for SN group, developed the implemented the code and the SN population along with their assignment to cosmoDC2 host galaxies, the cadence selection and contributed to the validation of SN, the planning and requirements for strong lensing injection, helped with validation of SN done by JWP, DS, RH, and SJ.
JRB contributed to production of the truth catalogs and to the software package GCRCatalogs.
DB contributed to the the image processing pipeline configuration, deployment and tuning at CC-IN2P3 and to the validation of the various data products.
KC contributed to the simulation and data processing workflows and Globus distribution portal.
JC worked on imSim development, image validation, image processing development and debugging, and calibration product generation.
JCT was responsible for the definition, implementation, and deployment of the SRS pipeline at CC-IN2P3.
AJC led the development of the LSST simulation tools and contributed to the initial definition of the DESC data challenges.
ADW developed the LSST DESC exposure checker and organized the DC2 visual inspection effort.
RD assisted in organization, planning and obtaining computing resources.
SFD helped design and implement the stellar and AGN variability models. He also implemented and maintained the interface between the cosmoDC2 simulations, the GalFast simulations, and ImSim.
SWD edited the note text.
EG contributed to the field location and dither design.
TG worked on the production of certain calibration products, and assisted with management of DESC NERSC resources.
SH is the HACC team lead; he contributed to the assessment of image generation computational requirements, co-led the management of DESC NERSC resources.
APH helped design and build the model of the galaxy-halo connection used to generate the cosmoDC2 extragalactic catalog. 
KH was responsible for the overall organization of the DC2 project, was involved in many aspects of the extragalactic catalog production, and contributed to the text of the note.
FH implemented the mechanism for making the LSST Science Pipelines available online and usable both at CC-IN2P3 and at NERSC, managed the CC-IN2P3 data processing infrastructure used by the image processing pipeline and was responsible for the prompt data transfer between CC-IN2P3 and NERSC.
RH worked on the coordination and testing of simulated SN inside DC2, draft reading and editing.
JH was a core member of the extragalactic catalog production team.
MJ contributed significant portions of code to both the GalSim and ImSim code bases for the purposes of generating the DC2 images.  He also contributed to the simulation design, especially decisions about which features should be included to achieve the desired goals of realism in the galaxy shapes for weak lensing science, while maintaining computational feasibility.
JBK was the main developer of the SL Sprinkler that inserted strongly lensed AGN into the instance catalogs and contributed the text of the paper relating to the SL Sprinkler.
HMK set up the web portal and managed the DESC software and data resources at NERSC. 
EK was one of the principal developers of the extragalactic catalog that was used as input to the image simulations and also worked on the validation of the DC2 object catalogs, as described in the DC2 survey paper.
DK led the development of the model underlying the extragalactic catalog.
KSK contributed to the conceptual design of the simulated survey including determining which electronic effects to simulate and by association which master calibration products to include.
FL contributed the model for the knots component included in galaxy light profiles, and the implementation of said model in CatSim and imSim.
PL made significant contributions to the development of the cosmoDC2 extragalactic catalog
CSL helped develop physical models of the CCD detectors, which allowed physically real simulations of tree rings and the brighter-fatter effect.
NL contributed to the generation of strongly lensed host galaxies of multiply lensed AGN and SNIa in the strong lensing systems sprinkled in the DDF.
EPL made contributions to the sky model in imSim. 
RHL contributed to the validation of the final data catalogs and provided support in using the LSST Science Pipelines.
RM organized analysis teams and synthesized input that factored into the overall DC2 design and validation, was engaged in the validation efforts, and edited the note text. 
YYM contributed to the generation, validation, and access of various DC2 data products, the preparation of public release, and text of this note.
PJM helped design the survey regions and cadences, provided high-level scientific oversight, and contributed to defining the strong lensing requirements.
JEM helped develop and validate the PSF simulation within imSim.
JWP contributed to the generation and documentation for the truth tables of strongly lensed SNe and AGN for the full DC2 effort.
JP contributes to write notebooks using Apache Spark to access and manipulate the DC2 data.
DJP implemented a model for LSST optical effects in imSim, assisted in the development of internal data access tools, and contributed to the visual validation of DC2 images.
JP implemented a system for running imSim on the UK computational grid and used it to perform parts of the image simulation runs in Europe.
SP contributed to the validation of various DC2 data products, and managed the Apache Spark tools at NERSC.
AP contributed to many aspects of the underlying extragalactic catalog and performed initial studies of using imSim in containers.
ESR contributed coverage mapping, processing QA for missing tracts, and galaxy color QA.
FJS participated in DC2 design phase, and production. Participated in catalog validation and matching between object and truth catalog.
SJS wrote the text for the photometric redshifts section.
TDU was involved in setting up the initial imSim simulations to scale them up  on thousands of nodes of Theta and supporting clusters at Argonne.
ASV was responsible for early generation of instance catalogs, implementing the Parsl workflow for imSim on NERSC and ALCF resources, and helping in initial validation of these outputs.
CWW carried out early planning for DC2, worked on development, testing and management of the imSim image simulation program, and tested the released data product format.
MPW implemented the code to add lensed host galaxies to the lensed AGN and lensed SNe in the DC2 code.
MWV co-led the Data Access Task Force, served as the Data Coordinator, and contributed to validation of the DC2 data products.
%

\clearpage
\phantomsection
\addcontentsline{toc}{section}{References}

\bibliographystyle{aasjournal}
\bibliography{ref}

\end{document}